\documentclass[aps,prd,a4paper,onecolumn,amsmath,showpacs,superscriptaddress,nofootinbib,preprintnumbers,notitlepage]{revtex4-1}
\usepackage{amssymb,amsmath}
\usepackage{graphicx}
\usepackage{color}
\usepackage{dcolumn}
\usepackage{pgfplots}
\usepackage{booktabs}
\usepackage{multirow}
\usepackage{dcolumn}
\usepackage{amsmath}
\usepackage{mathtools}
\usepackage{amsfonts}
\usepackage{amssymb}
\usepackage{epstopdf}
\usepackage{bm}
\usepackage{siunitx}
\usepackage{braket}
\usepackage{enumitem}
\usepackage{soul}
\usepackage{ulem}
\usepackage{color}
\usepackage{transparent}
\usepackage{pifont}
\usepackage[font={small}]{caption}
\newcommand\rf[1]{(\ref{eq:#1})}
\newcommand\lab[1]{\label{eq:#1}}

\newcommand\br{\begin{eqnarray}}
\newcommand\er{\end{eqnarray}}
\newcommand\be{\begin{equation}}
\newcommand\ee{\end{equation}}

\newcommand\lb{\lbrack}
\newcommand\rb{\rbrack}

\newcommand\bc{\begin{center}}
\newcommand\ec{\end{center}}

\newcommand\partder[2]{\frac{{\partial {#1}}}{{\partial {#2}}}}

\newcommand\eps{\epsilon}
\newcommand\vareps{\varepsilon}

\newcommand\G{\Gamma}

\renewcommand\k{\kappa}
\renewcommand\l{\lambda}

\newcommand\m{\mu}
\newcommand\n{\nu}

\renewcommand\P{\Phi}
\newcommand\pa{\partial}

\newcommand{\ct}[1]{\cite{#1}}
\newcommand{\bib}[1]{\bibitem{#1}}

\newcommand\PRD[3]{\textsl{Phys. Rev.} \textbf{D#1} (#2) #3}

\newcommand\PLB[3]{\textsl{Phys. Lett.} \textbf{#1B} (#2) #3}
\newcommand\CQG[3]{\textsl{Class. Quantum Grav.} \textbf{#1} (#2) #3}

\newcommand\IJMPD[3]{\textsl{Int. J. Mod. Phys.} \textbf{D#1} (#2) #3}

\newcommand\MPLA[3]{\textsl{Mod. Phys. Lett.} \textbf{A#1} (#2) #3}

\begin{document}
\title {Holomorphic General Coordinate Invariant Modified Measure Gravitational Theory.}

\author{Eduardo Guendelman}
\email{guendel@bgu.ac.il}
\affiliation{Department of Physics, Ben-Gurion University of the Negev, Beer-Sheva, Israel.\\}
\affiliation{Frankfurt Institute for Advanced Studies (FIAS),
Ruth-Moufang-Strasse 1, 60438 Frankfurt am Main, Germany.\\}
\affiliation{Bahamas Advanced Study Institute and Conferences, 
4A Ocean Heights, Hill View Circle, Stella Maris, Long Island, The Bahamas.
}
\begin{abstract}
Complexifying space time has many interesting applications, from the construction of higher dimensional unification,  to  provide a useful framework for quantum gravity and to better define some local symmetries that suffer singularities in real space time. In this context here spacetime is extended to complex spacetime and standard general coordinate invariance  is also extended  to complex holomorphic general coordinate transformations. 
This is possible by introducing a non Riemannian Measure of integration, which transforms avoiding non holomorphic behavior . Instead the measure transforms according to the inverse of the jacobian of the coordinate transformation and avoids the traditional square root of the determinant of the metric $\sqrt{-g}$. which is not globally holomorphic , or the determinant of the vierbein which is sensitive to the vierbein orientations and not invariant under local lorentz transformations with negative determinants. A contribution to the cosmological term appears as an integration constant in the equations of motion.
A proposed action for Finsler geometry, which involves $-g$ rather than $\sqrt{-g}$
will also constitute an example of a Holomorphic General Coordinate Invariant Modified Measure Gravitational Theory.

\end{abstract}
\maketitle
\section{Introduction}
\label{intro}

Complexified space time has attracted some attention in several somewhat disconnected contexts. For example Esposito and  others use complex space time as a way to represent 4 dimensional space through two complex coordinates \cite{Esposito}, This does not represent a modification of General Relativity, but rather introducing a new tool to study it,
Others propose higher dimensional complex spaces as a way to unify interactions
\cite{BLAHA} .

Here we will consider the replacement of the 4 dimensional space time by a 4 dimensional complex space time, as for example Ivanshuk \cite{Ivashchuk}, Moffat \cite{Moffat1}, \cite{Moffat2} , \cite{Moffat3}  , \cite{Moffat4}  and Chamsedinne \cite{Chamseddine} have considered . In  \cite{tHooft} the  simple change of a the manifold of real coordinates to an imaginary one was considered, However there is no direct correspondence between the transformation considered in \cite{tHooft}  and the holomorphic general coordinate transformations considered here, since in our case we would consider the transformation  coordinates  $x^\m \rightarrow ix^\m$  acting on the full complex space, not just the real coordinates coordinate transforming them into imaginary coordinates, but also, among other things,  imaginary coordinates transforming into real coordinates, so that the manifold would remain invariant, and we do not consider transforming one real manifold into another imaginary manifold as in  \cite{tHooft}.

Boundary conditions may be  real, at least in the classical theory. All kind of features, like symmetries become more transparent in the complexified theory and in the quantum theory, when tunneling and other effects are considered,  the complex space time becomes inevitable. Also,
another way complex space time could appear is from noncommutative space time\cite{quantumspacetime},
as discussed for example in in \cite{Chamseddine},  \cite{Subhash},  \cite{Moffat1}. 

We will see here how one can construct a complex, holomorphic or meromorphic theory of gravity where the space time is extended to complex space time and General coordinate invariant which has extended to holomorphic general coordinate transformations.  This is possible by introducing a non Riemannian Measure of integration, which transforms according to the jacobian of the coordinate transformation, Crucially, the non Riemannian Measures that we will use avoid the
square root of the determinant of the metric used in the standard formulation of general coordinate invariant theories, which is not holomorphic as we see first in our next section.


\section{ Non Holomorphic Structure of the Acion in General Relativity and similar theories with the Riemannian Measure}
\label{GR}
The action of GR, and other theories that use the standard Riemannian volume element  $ d^4 x {\sqrt{-g}}$  is of the form,
\be
S = \int d^4 x {\sqrt{-g}} L
\lab{GRL}
\ee
where $L$ is a generally coordinate invariant lagrangian.
Now notice the non holomorphic structure due to the appearance of  $\sqrt{-g}$, that under a general coordinate transformation, even when holomorphic,
$$ d^4 x \rightarrow Jd^4 x $$ , while  $\sqrt{-g} $ needs to be defined,
for example if $J$ is real and negative, we can define
$$ \sqrt{-g} \rightarrow  \mid J \mid ^{-1}\sqrt{-g}$$
where $J$ is the jacobian of the transformation and $ \mid J \mid$ is the absolute value of the transformation. Therefore $d^4 x {\sqrt{-g}} \rightarrow  \frac{J}{ \mid J \mid} d^4 x {\sqrt{-g}} $, so invariance is achieved only for $J = \mid J \mid$, that is if $J>0$, that is signed general coordinate transformations are problematic, and ill defined.

One could argue that when taking the square root of the determinant of the metric one may choose the negative solution when it suit us, but this would be an arbitrary procedure if no specific rule is given to choose the positive or the negative root. We choose instead to declare that $ \sqrt{-g} $ is always positive and replace it in the measure by something else whose sign is well defined. 

Recall that $0$ is a branch point of the square root function. Suppose 
$ w = \sqrt{z} $, and $z$ starts at $4$ and moves along a circle of radius $4$ in the complex plane centered at $0$. When we move in the complex circle from 4 we start with 2, but after we do the full circle we get $-2$, not 2. Obviously one of the definitions of the function will not leave the volume element invariant.

Of course we need to  define a measure that will transform holomorphically , the Riemannian measure, which makes use of the square root is not acceptable. Of course complex coordinate transformations will be even more problematic than just real and with a negative jacobian. 

\section{Metric Independent non-Riemannian Volume-Forms and Volume elements invariant under holomorphic  general coordinate transformations}
One can define a metric independent measure from a totally anti symmetric tensor gauge field, for example
\be
 \Phi (A) = \frac{1}{3!}\vareps^{\m\n\k\l} \pa_\m A_{\n\k\l} \quad ,
\lab{Phi}
\ee
where $A_{\n\k\l}$ is a three index totally antisymmetric tensor.
Then, under a complex holomorphic general coordinate transformation $$\Phi (A) \rightarrow J^{-1}\Phi (A) $$. . Therefore $d^4 x \Phi(A)  \rightarrow  d^4 x \Phi(A) $, so invariance is achieved  $J$ and we do not have to deal with how to define square roots, etc.
Of course the invariance of the volume element is only part of the discussion, we also have to discuss the lagrangian. It is worthwhile noticing that 3-index totally antisymmetric gauge fields have been used in other interesting applications, see for example \cite{Aurilia} 
 
\section{Theory using Metric Independent Non-Riemannian Volume-Forms, Non Holomorphic and Holomorphic cases}
\label{TMMT}

\subsection{Theory using Metric Independent non-Riemannian Volume-Forms, non Holomorphic cases}
\label{TMMT}
First we review our previous papers where we have considered  the  action 
of the general form involving two independent non-metric integration measure densities, \cite{TMT-orig-1}, \cite{TMT-orig-2} , \cite{TMT-orig-3} \cite{TMT-recent-1-a}, \cite{TMT-recent-1-b}, \cite{TMT-recent-1-c}, \cite{Comelli} , \cite{RCordero}, \cite{RCordero2},  \cite{DenitsaStaicova},  \cite{guendelmanKineticDEDM}, \cite{susy-break}, \cite{quintess}, \cite{curv}, \cite{ourquintessencewithEDE}, \cite{Gravityassisted}. 

Discussion of invariance of volume elements under general coordinate transformations that have jacobians that can be negative was discussed in  \cite{signed}, here we go beyond this, to the complex holomorphic case. 

For example
 generalizing the model analyzed in \ct{quintess} is given by 
\be
S = \int d^4 x\,\P_1 (A) \Bigl\lb R + L^{(1)} \Bigr\rb +  
\int d^4 x\,\P_2 (B) \Bigl\lb L^{(2)} + \eps R^2 + 
\frac{\P (H)}{\sqrt{-g}}\Bigr\rb \; .
\lab{TMMT1}
\ee

Here the following definitions  are used:

\begin{itemize}
\item
The quantities $\P_{1}(A)$ and $\P_2 (B)$ are two densities and these are  independent non-metric volume-forms defined in terms of field-strengths of two auxiliary 3-index antisymmetric
tensor gauge fields
\be
\P_1 (A) = \frac{1}{3!}\vareps^{\m\n\k\l} \pa_\m A_{\n\k\l} \quad ,\quad
\P_2 (B) = \frac{1}{3!}\vareps^{\m\n\k\l} \pa_\m B_{\n\k\l} \; .
\lab{Phi-1-2}
\ee
The density $\P (H)$ denotes  the dual field strength of a third auxiliary 3-index antisymmetric
tensor 
\be
\P (H) = \frac{1}{3!}\vareps^{\m\n\k\l} \pa_\m H_{\n\k\l} \; .
\lab{Phi-H}
\ee

\item
The scalar curvature $R = g^{\m\n} R_{\m\n}(\G)$ and the Ricci tensor $R_{\m\n}(\G)$ are defined in the first-order (Palatini) formalism, in which the affine
connection $\G^\m_{\n\l}$ is \textsl{a priori} independent of the metric $g_{\m\n}$.

\item
The two different Lagrangians $L^{(1,2)}$ correspond to two  matter field Lagrangians 
\end{itemize}
On the other hand, the variation of  \rf{TMMT1} w.r.t. auxiliary tensors 
$A_{\m\n\l}$, $B_{\m\n\l}$ and $H_{\m\n\l}$ becomes
\be
\pa_\m \Bigl\lb R + L^{(1)} \Bigr\rb = 0 \quad, \quad
\pa_\m \Bigl\lb L^{(2)} + \eps R^2 + \frac{\P (H)}{\sqrt{-g}}\Bigr\rb = 0 
\quad, \quad \pa_\m \Bigl(\frac{\P_2 (B)}{\sqrt{-g}}\Bigr) = 0 \; ,
\lab{A-B-H-eqs}
\ee
whose solutions are
\be
\frac{\P_2 (B)}{\sqrt{-g}} \equiv \chi_2 = {\rm const} \;\; ,\;\;
R + L^{(1)} = - M_1 = {\rm const} \;\; ,\;\; 
L^{(2)} + \eps R^2 + \frac{\P (H)}{\sqrt{-g}} = - M_2  = {\rm const} \; .
\lab{integr-const1}
\ee
Here the parameters $M_1$ and $M_2$ are arbitrary dimensionful and the quantity $\chi_2$ corresponds to an
arbitrary dimensionless integration constant. 

The resulting theory is called a Two Measure Theory, due to the presence of the Two measures  $\P_1 (A) $ and $\P_2 (A) $. But for the purpose of this paper this is two general, since we want to restrict to a theory that will give us ordinary General Relativity, and we want to keep the general coordinate invariance under complex holomorphic  general coordinate invariance. 

For obtaining GR dynamics, we can restrict to one measure, so let us take 

$$ \P_1 (A)= \P_2 (B) = \Omega $$

also can define the measure in terms of an additional set of four fields is introduced,  we express $ \Phi $ in terms of
four scalar fields
\be
\Omega = \frac{1}{3!}\vareps^{\m\n\k\l}\vareps^{abcd} \pa_\m\varphi_a \pa_\n\varphi_b \pa_\k\varphi_c \pa_\l\varphi_d \quad  
\lab{omega}
\ee
(one has to point out that in the earlier formulations of modified measures theories we used the 4 scalar field representation for the measure, see \cite{TMT-orig-1} , see also \cite{Pirogov} and \cite{Pirogov2} for more general theories using four scalars in four dimension, including possibly using them for defining measures of integration)
The mapping of the four scalars to the coordinates $x^\mu$ may be topologically non trivial.  Finally, we have to correct the equation
\be
\frac{\P_2 (B)}{\sqrt{-g}} \equiv \chi_2 = {\rm const}  .
\lab{eq.tobe corrected}
\ee
\subsection{Theory using Metric Independent non-Riemannian Volume-Forms, holomorphic cases}
 
For another equation that will be invariant under holomorphic general coordinate invariant transformations, 
we must avoid $\sqrt{-g}$, such an equation  which will be, 
\be
\frac{\Omega^2}{(-g)}\equiv \chi = K =  {\rm const}   .
\lab{corrected}
\ee
The resulting action that replaces \rf{Phi-1-2} is, 
\be
S = \int d^4 x\,\Omega \Bigl\lb R + L \Bigr\rb +  
\int d^4 x\,\Omega^2 \Bigl\lb 
\frac{\P (H)}{{(-g)}}\Bigr\rb \; .
\lab{simpleGCISIGNED}
\ee
the density $\P (H)$ remains defined eq. \rf{Phi-H}
so the integration obtained from the variation of the $H$ gauge field is  eq. \rf{corrected} now.
The solution of  eq. \rf{corrected} are
\be
\frac{\Omega}{ \sqrt{(-g)}} = \pm{\sqrt{K}}   .
\lab{solutions of corrected}
\ee
where the sign in \rf{solutions of corrected} will  be dynamically determined,
The measure $\Omega$ could have a small imaginary part. Of course  this imaginary part can be set to zero by initial conditions, but that is not mandatory.

Notice that although the theory has the complex holomophic invariance, a particular solution (which here means choosing between the plus or minus)  does not have to be, although the space of all solution is holomorphic invariant.  The restriction to some sign breaks the holomorphic invariance.  
\section{The determinant of the vierbein is a non invariant measure under signed Local Lorentz Transformations}
Another possibility for a measure that would transform like the the jacobian of the coordinate transformation, not the absolute value of the jacobian,  would be the determinant of the vierbein. This will destroy however (up to a sign) the invariance of the theory under signed local Lorentz transformation of the vierbeins. that is Local Lorentz transformations with negative determinants, so, it is not a solution, rather we trade one asymmetry for another.

\section{ The four scalars as integration manifold}
\label{invariantscalarintegrationmanifold}

Notice that using the volume element converts the the integration over coordinates in the action into integration over scalar fields, since $$\Phi d^4x = d\varphi_1 d\varphi_2 d\varphi_3 d\varphi_4$$. The scalars are complex as the original coordinates. The mapping of the coordinates to the scalars may not be one to one.

The scalar integration manifold existing in the four scalar field manifold is in fact completely unaffected by any holomorphic coordinate transformation taking place in the $x$ space. The lagrangian density is also a scalar not affected by any holomorphic coordinate transformation.


\section{Gravitational Equations of motion}
\label{Einstein}

We start by considering the equation that results from the variation of the degrees of freedom that define the measure $\Omega$,  that is the scalar fields $\varphi_a$, these are,
\be
A^{\m}_a \pa_\m (R + L  +2 \Omega \frac{\P (H)}{{(-g)}}) = 0
\lab{EINSTEIN}
\ee
where 
\be
A^{\m a} =\frac{1}{3!}\vareps^{\m\n\k\l}\vareps^{abcd}  \pa_\n\varphi_b \pa_\k\varphi_c \pa_\l\varphi_d \quad  
\lab{AMATRIX}
\ee
Notice that the determinant of $A^{\m a}$ is proportional to $\Omega^3 $, so if the measure is not vanishing, the matrix $A^{\m a}$ is non singular and therefore $\pa_\m (R + L  +2 \Omega \frac{\P (H)}{{(-g)}}) = 0 $, 
so that,
\be
R + L  +2 \Omega \frac{\P (H)}{{(-g)}} = M = constant
\lab{M}
\ee

The variation with respect to the metric $g^{\m\n}$, we obtain.
\be
\Omega (R_{\m\n}+ \frac{\pa L}{\pa g^{\m\n} } )  +   g_{\m\n} \Omega^2 \frac{\P (H)}{{(-g)}} = 0
\lab{munu}
\ee
solving $\Omega \frac{\P (H)}{{(-g)}}$  from \rf{M} and inserting into \rf{munu}, we obtain,
\be
R_{\m\n} - \frac{1}{2}g_{\m\n} R  + \frac{1}{2} M g_{\m\n} + \frac{\pa L}{\pa g^{\m\n} } - \frac{1}{2}g_{\m\n} L = 0
\lab{EINSTEINLIKEEQ}
\ee
which gives exactly the form of Einstein equation with the canonical energy momentum defined from $L$

\be
T_{\m\n} = g_{\m\n} L - 2 \partder{}{g^{\m\n}} L \; .
\lab{EM-tensor}
\ee

The equations of motion of the connection (in the first order formalism) implies that the connection is the Levi Civita connection.  L can describe a scalar field with the potential and the term $\frac{1}{2} M $ can be interpreted as a shift of the scalar field potential by a constant or a floating contribution to the cosmological constant. Notice that there no way to introduce an explicitly a cosmological constant term in the action.

One issue that should be addressed is that of the gauge fixing in the $\varphi_a$ space. Indeed, we notice that the only thing where these fields appear in the equations of motion is $\Omega$, but this quantity is invariant under volume preserving diffeomorphisms of the fields $\varphi_a$, $\varphi^\prime_a = \varphi^\prime_a(\varphi_a)$  which satisfy

\be \lab{VOLPRESDIFF}
\epsilon_{a_1 a_2 a_3 a_4}\frac{\partial{\varphi^\prime}_{b_1}}{\partial\varphi_{a_1}}\frac{\partial{\varphi^\prime}_{b_2}}{\partial\varphi_{a_2}}\frac{\partial{\varphi^\prime}_{b_3}}{\partial\varphi_{a_3}}\frac{\partial{\varphi^\prime}_{b_4}}{\partial\varphi_{a_4}} = \epsilon_{b_1 b_2 b_3 b_4}
\ee

so the study of the best gauge for the $\varphi_a$  fields for further comparison with the  $x^\m$ space could be a very important subject. Of course when we say that the mapping between the $\varphi_a$ and the $x^\m$ spaces, we want to exclude multi valuedness due to volume preserving diffeomorphisms of the fields $\varphi_a$, if for example different signs for $\Omega$ are associated to the same point in  $x^\m$ space, it is clear that there are at least two points  in $\varphi_a$ space associated to one point in  $x^\m$ space, and these two points in  the $\varphi_a$ are not related through a volume preserving diff. 

\section{Spontaneous Breaking of Holomorphic General coordinate Invariance in the Solutions}
The integration of the equations of motion can introduce some constants and then after this some choices of signs after the resolution of a quadratic equation. 

We have seen that the integration constant $M$ found in enters  and plays a role of a contribution to the cosmological constant . It has been introduced through equation  \ref{M}. This equation does not violate holomorphic general coordinate invariance, since  $M$  is a constant that does not transform under holomorphic  general coordinate invariance and the other side of that equation \ref{M} is also a scalar which is therefore also invariant, no spontaneous breakdown of holomorphic general coordinate general coordinate invariance .

The quadratic  eq. \rf{corrected}  satisfies holomorphic general coordinate invariance but its  solutions, when we restrict to a particular sign \rf{solutions of corrected} do not. 
To show this let us consider for example a real but signed general coordinate transformation, where as we have seen the two measures we are dividing transform with different signs, since $$ \sqrt{-g} \rightarrow  \mid J \mid ^{-1}\sqrt{-g}$$ 
$$ \Omega \rightarrow J ^{-1} \Omega $$
which means that under such transformations the solution with plus transforms into the solution with minus and vice versa. 
\be
\frac{\Omega}{ \sqrt{(-g)}}  \rightarrow  \frac{\mid J \mid }{J}  \frac{\Omega}{ \sqrt{(-g)}}
\lab{changing sign}
\ee
and since  $J$ is negative we have a change of sign in theratio of the two measures.
\section{ Slightly modified Finsler geometry Action as an example of a Holomorphic General Coordinate Invariant Modified Measure  Theory}

In a recent paper there was a proposal for an action for Finsler geometry \cite{Pfeifer}. The idea is to consider coupling of the determinant of the metric but not to the volume of space time, but rather. To the phase space volume, so the tangent space enters in the volume as well, see eq (22) in \cite{Pfeifer}. The authors suggest coupling this phase space volume to $\mid g \mid $, the absolute value of the determinant of the metric, but  for the holomorphic version of this action we would just couple to $-g$, i.e. , minus the determinant of the metric, which for Minkowski signature real solutions is also positive, but has a different continuation to complex space that respects holomorphic behavior.

\section{Discussion}
We have discussed how general coordinate invariance is extended to complex holomorphic general coordinate transformations.



A local realization of complex holomorphic general coordinate transformation invariance is possible by introducing a non Riemannian Measure of integration, not involving square roots as it  is the case with $\sqrt{-g}$.

The modified measure theory we have described here does this and the sign in \rf{solutions of corrected} will  be dynamically determined.

One could also go beyond General Relativity like theories, and generalize the modified measure theory \rf{TMMT1} in the following way that would respect complex holomorphic general coordinate invariance, 
\be
S = \int d^4 x\,\P_1 (A) \Bigl\lb R + L^{(1)} \Bigr\rb +  
\int d^4 x\,\P_2 (B) \Bigl\lb L^{(2)} + \eps R^2 + 
\P_2 (B)\frac{\P (H)}{(-g)}\Bigr\rb \; .
\lab{TMMTsigned}
\ee
this will be now also a signed general coordinate invariant theory. One can again explore a representation of modified measures in terms of 4 scalar fields, etc. This case will be more involved, requiring solving for two measures, defining an Einstein frame, etc, following for example the work done in \cite{quintess}. This will be done in a future publication.

Then there are questions related to the quantum theory that should be related to the complexified general coordinate invariant theory. In particular the question of a cosmological constant seems of particular relevance, since the cosmological constant cannot be added as a contribution to the action, can appear only as an integration constant of the equations of motion, so it appears as an infrared problem, not an ultraviolet problem.

Another subject to study:  since the cosmological constant appears as an integration constant, one could as if it could have a very small imaginary condition, likewise whether the measure also could have a small imaginary part and what would be the consequences. Of course all these imaginary parts can be set to zero by initial conditions, but that is not mandatory.

 One can also formulate the complex  holomorphic reparametrization invariace for theories of extended objects using an appropriate version of the modified measure formulation of strings and branes done in  \cite{mstring},  \cite{nishino-rajpoot}, \cite{mstringspectrum}, \cite{mstringbranes},  \cite{stringsandantistrings}.
The study of the holomorphic actions for Finsler geometry could be also worthwhile.
 Finally one could explore how these possibilities here explored for Modified Measures theories also extend to  Causal Fermion Systems, since there are many similar features between these approaches \cite{MMTcfs}.


\begin{acknowledgements}
I want to thank Emil Nissimov and Svetlana Pacheva for interesting conversations, organizers of the BENASQUE Cosmology and the Quantum Vacuum for inviting me to present a talk on the subject of this paper and very interesting conversations with the participant,  FQXi  for great financial support for work on this project at BASIC in Ocean Heights, Stella Maris, Long Island,  Bahamas, CosmoVerse  COST Action CA21136 and  COST Action CA18108 - Quantum gravity phenomenology in the multi-messenger approach for financial support.
\end{acknowledgements}

\end{document}